\documentstyle[aps,prb,epsf, multicol]{revtex}
\tighten
\begin{document}
\draft
\newcommand{\bn}{{\bf n}}
\newcommand{\bp}{{\bf p}}
\newcommand{\br}{{\bf r}}
\newcommand{\bq}{{\bf q}}
\newcommand{\bj}{{\bf j}}
\newcommand{\bE}{{\bf E}}
\newcommand{\eps}{\varepsilon}
\newcommand{\la}{\langle}
\newcommand{\ra}{\rangle}
\newcommand{\cK}{{\cal K}}
\newcommand{\cD}{{\cal D}}
\newcommand{\sign}{\mathop{\rm sign}\nolimits}
\newcommand{\mybeginwide}{
    \end{multicols}\widetext
    \vspace*{-0.2truein}\noindent
    \hrulefill\hspace*{3.6truein}
}
\newcommand{\myendwide}{
    \hspace*{3.6truein}\noindent\hrulefill
    \begin{multicols}{2}\narrowtext\noindent
}
\newcommand{\be}{\begin{equation}}
\newcommand{\ee}{\end{equation}}

\title{
  Frequency Scales for Current Statistics of Mesoscopic Conductors
}

\author{
   K. E. Nagaev,$^{1,2}$ S. Pilgram,$^2$ and M. B\"uttiker$^2$
}
\address{
  $^1$Institute of Radioengineering and Electronics,
  Russian Academy of Sciences, Mokhovaya ulica 11, 125009 Moscow,
  Russia\\
  $^2$D\'epartement de Physique Th\'eorique, Universit\`e de
  Gen\'eve, CH-1211, Gen\`eve 4, Switzerland\\
}

\date\today
\maketitle

\begin{abstract}
We calculate the third cumulant of current in a chaotic cavity with
contacts of arbitrary transparency as a function of frequency. Its
frequency dependence drastically differs from that of the conventional
noise. In addition to a dispersion at the inverse $RC$ time
characteristic of charge relaxation, it has a low-frequency dispersion
at the inverse dwell time of electrons in the cavity. This effect is
suppressed if both contacts have either large or small transparencies.
\end{abstract}

\begin{multicols}{2}
\narrowtext
\vspace{1cm}
What is the characteristic time scale of the dynamics of electrical transport in
normal-metal mesoscopic systems if their size is larger than the screening length? The
most reasonable answer is that this dynamics is governed by the $RC$ time of the system,
which describes the relaxation of piled up charge. It is this time that characterizes
the admittance of a mesoscopic capacitor,\cite{Buttiker-93} the  impedance of a
diffusive contact,\cite{Naveh-99} current noise in diffusive contacts,\cite{Nagaev-98}
and charge noise in quantum point contacts and chaotic
cavities.\cite{Pedersen-98,Blanter-00a} The free motion of electrons is superseded by
the effects of charge screening and therefore the dwell time of a free electron is not
seen in these quantities. The only known exception is the frequency dispersion of the
weak-localization correction to the conductance.\cite{Anderson-79,Brouwer-97} Because
typical charge-relaxation times are very short for good conductors, the experimental
observation of the dispersion of current noises is difficult.\cite{Schoelkopf-97}

Very recently, the third cumulant of current has been
measured for tunnel junctions.\cite{Reulet-03}
In this case, the frequency dispersion is due only to the
measurement circuit.\cite{Beenakker-03}
In contrast to tunnel junctions, chaotic
cavities have internal dynamics. It is
the purpose of this paper to show that mesoscopic systems with
such dynamics may exhibit an additional low-frequency dispersion
in the third cumulant of current, which is absent for the first and
second cumulants.
Physically, this dispersion is due to slow, charge-neutral
fluctuations of the distribution function
in the interior of the system. Such charge-neutral
fluctuations are similar to fluctuations of the
effective electron temperature at a constant chemical potential.
They do not contribute directly to the
measurable current, but they modulate the intensity of noise and hence
contribute to higher cumulants of current.\cite{Nagaev-02a}

The zero-frequency electric noise in open chaotic cavities has been calculated by
Jalabert et al.\cite{Jalabert-94} using random-matrix theory. More recently, these
expressions were derived semiclassically by Blanter and Sukhorukov.\cite{Blanter-00b}
Shot-noise measurements on chaotic cavities were performed by Oberholzer et
al.\cite{Oberholzer-01} The third and fourth cumulants of current were obtained by
Blanter et al.\cite{Blanter-01} for open cavities and in Ref.\cite{Nagaev-02b} for
cavities with contacts of arbitrary transparency. Recently, the statistics of charge in
a cavity was analyzed by two of us.\cite{Pilgram-03}
\begin{figure}[htb]
\epsfxsize8cm
\centerline{\epsffile{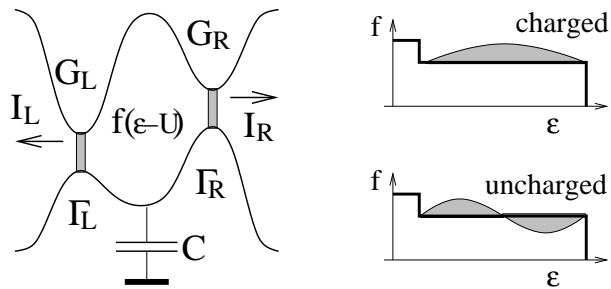}}
\vspace{5mm}
\caption{
{\bf Left panel:} A chaotic cavity
is biased from two leads and capacitively coupled to a gate.
{\bf Right panel:} Charged fluctuations of the distribution
function $f$ decay at the short RC time and contribute to the current
noise. Uncharged fluctuations decay at a much longer time $\tau_D$ and do not
contribute directly to the current noise but result in a low-frequency
dispersion of the third cumulant of current.
}
\label{Geometry}
\end{figure}
\noindent

The system we investigate is shown in Fig.\ \ref{Geometry}. A chaotic cavity consists of
a metallic mesoscopic conductor of irregular shape connected to leads $L,R$ via quantum
point contacts with conductances $G_{L,R}$ and transparencies $\Gamma_{L,R}$. We assume
that $G_{L,R} \gg e^2/h$ and the bias is much larger than the temperature and
frequencies. The dwell time of electrons in the cavity is given by $\tau_D = e^2 R_Q
N_F$, where $R_Q = (G_L + G_R)^{-1}$ is the charge-relaxation
resistance\cite{Brouwer-97} and $N_F$ is the density of states in the cavity, which is
assumed to be continuous and constant. We assume that the dwell time is short as
compared to inelastic scattering times.  Then the mean occupation function may be
written as a weighted average $f_0(\epsilon) = R_Q(G_L f_L(\epsilon) + G_R
f_R(\epsilon))$ of the Fermi occupation factors $f_{L,R}$ of the leads which are step
functions at zero temperature. In the absence of electrostatic interaction, fluctuations
around $f_0$ would relax on the time scale $\tau_D$. However, due to screening there is
a second shorter time scale $\tau_{Q}$, which describes fluctuations of the charge in
the cavity. It is given by $\tau_{Q} = R_Q C_{\mu}$, where $C_{\mu}^{-1} = C^{-1} +
(e^2N_F)^{-1}$ is the electrochemical capacitance\cite{Buttiker-93} and $C$ is the
geometric capacitance.

Our semiclassical calculations are based on a large separation between the time scales
describing the fast fluctuations of current in isolated contacts and slow fluctuations
of the electron distribution in the cavity.\cite{Nagaev-02b} This allows us to consider
the contacts as independent generators of white noise, whose intensity is determined by
the instantaneous distribution function of electrons in the cavity. Based on this
time-scale separation, a recursive expansion of higher cumulants of semiclassical
quantities in terms of their lower-order cumulants was
developed.\cite{Nagaev-02a,Nagaev-02b} Very recently, such recursive relations were
obtained as a saddle-point expansion of a stochastic path integral.\cite{Pilgram-02}
Here we follow the approach of Ref.\cite{Pilgram-02}

Our aim is to obtain the statistics of time-dependent fluctuations of
the current $I_L(t)$ flowing through the left contact of the
cavity. In general, the probability ${\cal P}[A(t)]{\cal D}A$ of a
time-dependent
stochastic variable $A(t)$ ($t\in [0,T]$) is described by the
characteristic functional $S[\chi_A]$ evaluated with
an imaginary field
\begin{eqnarray}
\label{Characteristic Functional}
{\cal P}[A(t)] = \int {\cal D}\chi_A
e^{-i\int_0^Tdt A\chi_A + S[\chi_A\mapsto i\chi_A]}.
\end{eqnarray}
Functional derivatives $\delta^n /(\delta \chi_A(t_1) \dots \delta
\chi_A(t_n))$ of the characteristic functional $S[\chi_A]$ yield
irreducible correlation functions
$\langle A(t_1)\dots A(t_n)\rangle$.

To obtain the characteristic functional $S[\chi_L]$ describing
the fluctuations of $I_L(t)$ we proceed in two steps.
First, we consider the point contacts
($i=L,R$) as sources of white noise that depend on two common
time-dependent parameters, the electron occupation function
of the cavity $f(\epsilon-eU)$ and the electrostatic
potential of the cavity $U$. Their characteristic
functionals are given by
$S_i[\tilde{\chi}_{i,\epsilon}] =
\int dt \int d\epsilon H_i(\tilde{\chi}_{i,\epsilon})$
with
\begin{eqnarray}
\label{Quantum Sources}
H_i =
 \frac{1}{2}\langle \tilde{I}_i^2 \rangle_{\epsilon}
\tilde{\chi}_{i,\epsilon}^2
+ \frac{1}{6}\langle \tilde{I}_i^3 \rangle_{\epsilon}
\tilde{\chi}_{i,\epsilon}^3
+ \dots.
\end{eqnarray}
The correlators $\langle \tilde{I}_i^n \rangle_{\epsilon}$
must be taken from a quantum-mechanical
calculation\cite{Levitov-93,Muzykantskii-94}
$$
 \langle
  \tilde{I}_i^n
 \rangle_{\epsilon}
 =
 \frac{G_i}{\Gamma_i}
 \frac{\partial^n}{\partial \chi^n}
 \ln
 \bigl\{
   1
   +
   \Gamma_i f_i(\epsilon)
   [1-f(\epsilon-eU)]
   (
    e^{-e\chi}-1)
$$ \begin{equation}
   +
   \Gamma_i
   f(\epsilon-eU)
   [1-f_i(\epsilon)]
   (
    e^{e\chi}-1
   )
 \bigr\}|_{\chi=0}.
 \label{Point Contacts}
\end{equation}
In a second step, we take into account that the occupation function $f(\epsilon-eU)$ and
the potential $U$ are not free parameters but are fixed by the kinetic equation
$$
 \left(
  \frac{d}{dt}
  +
  \frac{1}{\tau_D}
 \right)
 \delta f(\epsilon)
 =
  \frac{dU}{dt}
 \frac{ \partial f}{\partial U }
 -
 \frac{1}{N_F}
 \left(
  \tilde I_{L,\epsilon}
  +
  \tilde I_{R,\epsilon}
 \right)
$$
and the charge-conservation law
\begin{equation}
 C\frac{dU}{dt}
 =
 -
 \int d\epsilon
 \left[
  \frac{G}{e}
  \delta f(\epsilon)
  +
  \tilde I_{L,\epsilon}
  +
  \tilde I_{R,\epsilon}
 \right].
\label{Conservation Laws}
\end{equation}
We introduced the fluctuating
part of the occupation function $\delta f(\epsilon) =
f(\epsilon-eU) - f_0(\epsilon)$.
The two conservation laws are expressed by path integrals over Lagrange
multipliers $\lambda_{\epsilon},\xi$ and integrated over the
fluctuations of occupation function and potential to obtain the
following result for the generating functional
$$
 e^{S[i\chi_L]} = \int {\cal D}\lambda_{\epsilon} {\cal D}\xi
 {\cal D} f {\cal }
$$ \begin{equation}
 \times\exp\left\{S_L[i\lambda_{\epsilon}+i\xi-i\chi_L]
 + S_R[i\lambda_{\epsilon}+i\xi]  + S_C\right\},
 \label{Effective Generator}
\end{equation}
where the conservation laws are expressed by the following dynamical
action
$$
 S_C
 =
 -i\int_0^T dt
 \Bigl\{
  \xi
  \left[
   C \dot{U}
   +
   \int d\epsilon
   \frac{G}{e}\delta f(\epsilon)
  \right]
$$ $$
 \left.
  +
  \int d\epsilon
  \lambda_{\epsilon}
  \left[
   N_F
   \left(
    \frac{d}{dt}
    +
    \frac{1}{\tau_D}
   \right)
   \delta f
   -
   N_F
   \frac{dU}{dt}
   \frac{ \partial f}{\partial U}
  \right]
 \right.
 \Bigl.
$$ \begin{equation}
  -
  G_L \chi_L
  \int d\epsilon
  \left[
   f(\epsilon-eU) - f_L(\epsilon)
  \right]
 \Bigr\}.
 \label{Effective Action}
\end{equation}
In the semiclassical regime, this path integral may be evaluated in the saddle-point
approximation.\cite{Pilgram-02} The saddle point equations are nonlinear differential
equations for the four fields $f(\epsilon-eU), \delta U,\lambda_{\epsilon},\xi$ that
contain the external statistical field $\chi_L$ in inhomogeneous source terms. They
describe the non-linear response of internal fields to these sources. In this
publication, we are interested in the second and third order correlation functions of
the current $I_L$. For this purpose it is sufficient to expand the sum
$\eta=\lambda_{\epsilon}+\xi = \eta_1 + \eta_2/2 + \dots$ up to second order in the
external field ($\eta_n \propto \chi_L^n$). The result can then be substituted into the
action (\ref{Effective Generator}) to obtain the frequency dispersion of the correlation
functions.

We first briefly discuss the frequency dependence of current noise
to connect our theory to earlier results.\cite{Pedersen-98} To this
end we collect the
second order contributions of Eq. (\ref{Effective Generator})
and eliminate time derivatives using the saddle point equations
\begin{eqnarray}
 \label{Noise Action}
 S_2[\chi_L]
 =
 \frac{1}{2}
 \int dt d\epsilon
 \left[
  \langle \tilde{I}_L^2 \rangle_0
  (\eta_1 - \chi_L)^2
  +
  \langle \tilde{I}_R^2 \rangle_0
  \eta_1^2
 \right].
\end{eqnarray}
The zero subscript indicates that the correlators
$\langle \tilde{I}_i^n \rangle_{\epsilon}$
are evaluated at
$f(\epsilon-eU)=f_0(\epsilon)$.
We choose a large time interval $[0,T]$ and neglect any transient
effects. It is then most convenient to work completely in Fourier
space. The linear response of the internal fields
$(1+i\omega\tau_Q)\eta_1 = G_L R_Q \chi_L$ is entirely
governed by the RC time, therefore
the correlation function
$\langle I_L(\omega)I_L(\omega') \rangle =
2\pi\delta(\omega+\omega') P_2$
takes a simple form
\begin{eqnarray}
 \label{Noise Spectrum}
 P_2
 =
 |Z_Q(\omega)|^2
 \int d\epsilon
 \left[
  \langle \tilde{I}_L^2 \rangle_0
  |\tilde{G}_R(\omega)|^2
  +
  \langle \tilde{I}_R^2 \rangle_0
  G_L^2
 \right]
\end{eqnarray}
where we introduced $Z_Q(\omega) = (G_L+G_R+i\omega C_{\mu})^{-1}$ and
$\tilde{G}_R(\omega) = G_R + i\omega C_{\mu}$. At low frequencies, the
current
\begin{figure}[htb]
\epsfxsize7cm \centerline{\epsffile{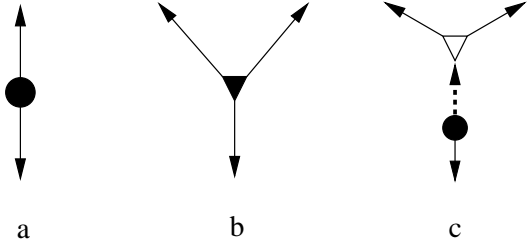}} \vspace{5mm} \caption{ Cascade expansion
for the third cumulant of current. (a) Second cumulant, (b) minimal-correlation third
cumulant, and (c) cascade correction to the third cumulant. The external ends correspond
to current fluctuations at different moments of time and the dashed line, to
fluctuations of the distribution function. The full circle and triangle correspond to
the second and third cumulants of white noise generated at the contacts. The white
triangle corresponds to the functional derivative of the second cumulant. }
\label{Diagrams}
\end{figure}
\noindent
noise shows correlations between left and right point contact. At high frequencies,
these correlations disappear and we observe the bare noise of the left point contact.
The transition frequency is given by the $RC$ time. To complete this result, it remains
to evaluate the bare noise correlators using Eq. (\ref{Point Contacts}).

We now turn to the much more complicated frequency dependence of the
third order correlator. For the action (\ref{Effective Action}) we
find
\begin{eqnarray}
 \label{Third Order Action}
 &&S_3[\chi_L]
 =
 \frac{1}{6}
 \int dt d\epsilon
 \left\{
  \langle
   \tilde{I}_L^3
  \rangle_0
  (\eta_1 - \chi_L)^3
  +
  \langle
   \tilde{I}_R^3
  \rangle_0
  \eta_1^3
 \right.
 \nonumber\\
 &&\qquad
 \left.
  +
  3
  \left[
   \langle
    \tilde{I}_L^2
   \rangle_0
   (\eta_1 - \chi_L)
   +
   \langle
    \tilde{I}_R^2
   \rangle_0
   \eta_1
  \right]\eta_2
 \right\}.
\end{eqnarray}
This correlator contains two contributions presented by diagrams in Fig. \ref{Diagrams}.
The first one represented by diagram $b$ is the minimal correlation
result\cite{Blanter-00b} that depends only on the $RC$ time. The second is represented
by diagram $c$ and gives the cascade correction,\cite{Nagaev-02a,Nagaev-02b} which
contains the low-frequency dispersion. The equations of motion are of the form
$$
 (1+i\omega\tau_D)\lambda_2
 =
 R_Q A_L
 (\eta_1 - \chi_L)^2
 +
 R_Q A_R
 \eta_1^2
 -
 \xi_2,
$$ \begin{equation}
 (1+i\omega\tau_Q)\xi_2
 =
 R_Q B_L(\eta_1 - \chi_L)^2
 +
 R_Q B_R\eta_1^2
 \label{Second Order Response}
\end{equation}
and depend on both time constants
$\tau_Q$ and $\tau_D$ that describe the decay of charged and
charge neutral fluctuations of the occupation function
$f(\epsilon-eU)$. In turn,  these fluctuations
act back on the current noise. We used here a notation
$A_i = \partial\la\tilde I_i^2\ra_0/\partial f$, and $B_i = (C_{\mu}/C)
\int d\epsilon (-\partial f_0/\partial \epsilon) A_i$. It is now
straightforward to substitute the
second order response (\ref{Second Order Response}) into
the action (\ref{Third Order Action}). Functional derivatives
with respect to the external field $\chi_L$ then yield the irreducible
part of the third order correlation function
$\langle I_L(\omega_1)I_L(\omega_2)I_L(\omega_3) \rangle =
2\pi\delta(\omega_1+\omega_2+\omega_3)P_3$.
The same results are obtained by using a recursive diagrammatic
expansion of the third cumulant.\cite{Nagaev-02a,Nagaev-02b}
\begin{figure}[htb]
\epsfxsize8cm \centerline{\epsffile{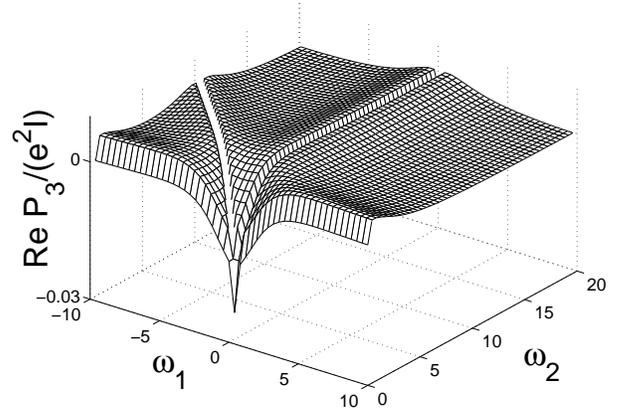}}
\vspace{5mm}
\caption{ A 3D plot of ${\rm
Re}\,P_3(\omega_1, \omega_2)$ for $G_L/G_R=1/2$, $\Gamma_L = \Gamma_R = 3/4$, $\tau_Q =
1/3$, and $\tau_D=10$ (dimensionless units). } \label{3D}
\end{figure}

In what follows, we will be interested in the most typical case where the dwell time
$\tau_D$ is much larger than the charge relaxation time $\tau_Q = C_{\mu}R_Q$. Even for
this case the general expression for $P_3$ is too long and we present here only its
limiting values. In the low-frequency limit $\omega_i \ll (C_{\mu}R_Q)^{-1}$, $P_3$ is
of the form\cite{complex}
$$
 P_3(\omega_1, \omega_2)
 =
 e^2 I
 \Biggl\{
  3G_L G_R
  \frac
  {
   \left[
    (1 - \Gamma_R) G_L^2
    -
    (1 - \Gamma_L) G_R^2
   \right]^2
  }{
   (G_L + G_R)^6
  }
$$ $$
  -
  2
  \frac
  {
   \Gamma_R^2 G_L^5
   +
   \Gamma_L^2 G_R^5
  }{
   (G_L + G_R)^5
  }
  +
  3
  \frac
  {
   \Gamma_R G_L^4
   +
   \Gamma_L G_R^4
  }{
   (G_L + G_R)^4
  }
  -
  \frac
  {
   G_L^3 + G_R^3
  }{
   (G_L + G_R)^3
  }
$$ $$
  -
  G_L G_R
  \frac
  {
   \left[
    (1 - \Gamma_R) G_L^2
    -
    (1 - \Gamma_L) G_R^2
   \right]
   \left(
    \Gamma_R G_L^2
    -
    \Gamma_L G_R^2
   \right)
  }{
   (G_L + G_R)^6
  }
$$ \begin{equation}
  \times
  \left[
   \frac{1}{ 1 + i\omega_1\tau_D }
   +
   \frac{1}{ 1 + i\omega_2\tau_D }
   +
   \frac{1}
   {
    1 - i(\omega_1 + \omega_2)\tau_D
   }
  \right]
 \Biggr\},
 \label{S-low}
\end{equation}
which in general suggests a strong dispersion of the third cumulant of noise at
$\omega_{1,2} \sim 1/\tau_D$. This dispersion vanishes for symmetric cavities and
cavities with two tunnel or two ballistic contacts. For a cavity with two tunnel
contacts, the white-noise sources (\ref{Point Contacts}) are linear functionals of the
distribution function and hence are not affected by charge-neutral fluctuations. For the
case of two ballistic contacts, Eq. (\ref{Quantum Sources}) depends only on $f$ and not
on $f_i$. Then the low-frequency dispersion does not show up either because fluctuations
of current and distribution function are uncorrelated due to the symmetry of $H_i$.
Note that Eq. (\ref{S-low}) is symmetric with respect to indices $L$ and $R$ because
pile-up of charge in the cavity is forbidden at low frequencies.

In general, $P_3(\omega_1, \omega_2)$ exhibits a complicated behavior (see Fig.
\ref{3D}). It has also a  dispersion at the inverse $RC$
\begin{figure}[htb]
\epsfxsize8cm \centerline{\epsffile{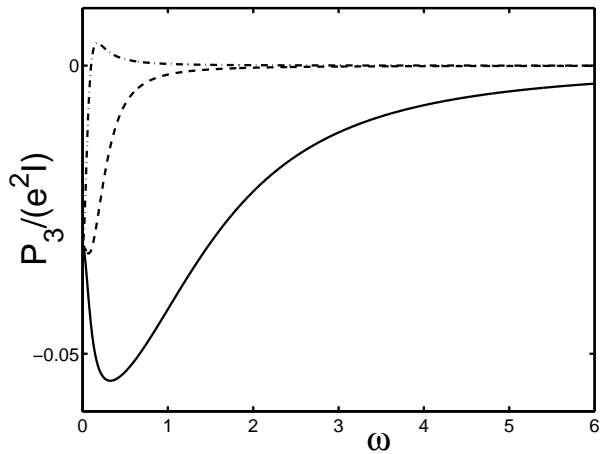}} \caption{$P_3(\omega, 0)$ as a function of
$\omega$ for $G_L/G_R=1$, $\Gamma_L=1$, $\Gamma_R=0$, and $\tau_D = 10$ (dimensionless
units). The solid, dashed, and dash-dot curves correspond to $\tau_Q =1/2$,  $\tau_Q
=3$, and to the case of weak electrostatic coupling $\tau_Q = \tau_D$ ($C=\infty$). }
\label{2D}
\end{figure}
\noindent
time and additional peculiarities at the scale $\tau_D^{-1}$ if one of the frequencies
or their sum tends to zero. The shape of $P_3(\omega_1, \omega_2)$ essentially depends
on the parameters of the contacts. In particular, for a cavity with one tunnel and one
ballistic contact with equal conductances $G_L = G_R = G$ it exhibits a non-monotonic
behavior as one goes from $\omega_1 = \omega_2 = 0$ to high frequencies. A relatively
simple analytical expression for this case may be obtained if $\tau_D \gg \tau_Q$ and
one of the frequencies is zero:
\begin{equation}
 P_3(\omega, 0)
 =
 -\frac{1}{32}
 e^2 I
 \frac
 {
  1
  +
  2\tau_D^2
  \omega^2
  +
  \tau_D^2\tau_Q^2
  \omega^4
 }{
  (1 + \omega^2\tau_D^2)
  (1 + \omega^2\tau_Q^2)^2
 }.
 \label{g10}
\end{equation}
%
The $P_3(\omega, 0)$ curve shows a clear minimum at $\omega \sim (\tau_D \tau_Q)^{-1/2}$
and the amplitude of its variation tends to $P_3(0, 0)$ as $\tau_Q/\tau_D \to 0$ (see
Fig. \ref{2D}).

The fluctuations of the current in the left-hand contact of the cavity can be measured
as fluctuations of voltage across a small resistor attached to it. Based on the
parameters of a chaotic cavity used in shot-noise experiments,\cite{Oberholzer-01} our
estimates give an inverse dwell time of the order of 10 GHz, which is well within the
experimental range for measuring the frequency dependence of noise.\cite{Schoelkopf-97}
We also believe that similar low-frequency dispersion of the third cumulant may be
observed in other semiclassical systems like diffusive wires, where the dwell time is of
the same order or larger. Hence our results are of direct experimental interest.

In summary, we have shown that the third cumulant of current in mesoscopic systems may
exhibit a strong dispersion at frequencies much smaller than the charge-relaxation time
of the system. This effect has a purely classical origin and the variations of the
cumulant may be of the order of its zero-frequency value even if the number of quantum
channels in the system is large.

This work was supported by the Swiss National Science Foundation, the program for
Materials and Novel Electronic Properties, the Russian Foundation for Basic Research,
Grant No. 01-02-17220, INTAS (project 0014, open call 2001), and by the Russian Science
Support Foundation.

\end{multicols}

\begin{thebibliography}{99}
\bibitem{complex}
$P_3(\omega_1, \omega_2)$ is complex if neither of the frequencies $\omega_1$,
$\omega_2$, and $\omega_1 + \omega_2$ is zero.
\end{thebibliography}
\end{document}